\documentclass[]{article}
\usepackage[margin=1in]{geometry}
\usepackage{times}
\usepackage{titlesec}
\usepackage{hyperref}
\usepackage{graphicx}
\usepackage{float}
\usepackage{amsmath}
\usepackage{caption}
\usepackage{fancyhdr}
\usepackage{multicol}
\usepackage{tcolorbox}
\usepackage{fancyvrb}
\usepackage{caption} 
\usepackage{listings}
\usepackage{cleveref}
\usepackage{xcolor}

\lstdefinestyle{Python}{
    language=Python,
    basicstyle=\ttfamily\footnotesize, 
    commentstyle=\color{gray}, 
    keywordstyle=\color{blue}\bfseries, 
    stringstyle=\color{red}, 
    showstringspaces=false, 
    tabsize=4, 
    frame=single, 
    captionpos=b, 
    numbers=left, 
    numberstyle=\tiny\color{gray}, 
    breaklines=true, 
    title=\lstname 
}

\definecolor{codebg}{RGB}{248,248,248}

\titleformat{\section}{\large\bfseries}{\thesection}{1em}{}
\titleformat{\subsection}{\normalsize\bfseries}{\thesubsection}{1em}{}

\hypersetup{
    pdftitle={Modeling Sparse and Bursty Vulnerability Sightings: Forecasting Under Data Constraints},
    pdfauthor={Cédric Bonhomme, Alexandre Dulaunoy},
    pdfsubject={Modeling Sparse and Bursty Vulnerability Sightings: Forecasting Under Data Constraints},
    pdfkeywords={CVE, Vulnerability, Forecast, Decision, SARIMAX, Poisson},
    colorlinks=true, linkcolor=blue, citecolor=blue, urlcolor=blue
}

\title{\textbf{Modeling Sparse and Bursty Vulnerability Sightings: Forecasting Under Data Constraints}}
\author{
    Cédric Bonhomme \\
    \textit{Computer Incident Response Center Luxembourg} \\
    \href{mailto:cedric.bonhomme@circl.lu}{cedric.bonhomme@circl.lu} [\href{https://openpgp.circl.lu/pks/lookup?op=get\&search=0xA1CB94DE57B7A70D}{57B7 A70D}]
    \and
    Alexandre Dulaunoy \\
    \textit{Computer Incident Response Center Luxembourg} \\
    \href{mailto:alexandre.dulaunoy@circl.lu}{alexandre.dulaunoy@circl.lu} [\href{https://pgp.circl.lu/pks/lookup?op=get\&search=0x3b12dcc282fa29312f5b709a09e2cd4944e6cbcd}{44E6 CBCD}]
}
\date{2025-11-05}

\begin{document}
\maketitle

\begin{abstract}
Understanding and anticipating the evolution of vulnerability-related activities is a key challenge in modern cyber threat intelligence. In this work, we investigate the feasibility of forecasting vulnerability sightings—that is, observable events reflecting public attention or exploitation potential, such as the appearance of proof-of-concept (PoC) code, detection templates, or discussions across the Fediverse and other platforms. Building upon our previous contribution, VLAI (\cite{bonhomme2025vlairobertabasedmodelautomated}), a transformer-based model that predicts vulnerability severity directly from textual descriptions, we explore how severity information can be incorporated as an explanatory factor in time-series forecasting models.

Our experiments evaluate multiple statistical and probabilistic approaches to short-term forecasting of sightings per vulnerability. We first assess SARIMAX models, with and without log-transformations and exogenous variables derived from VLAI severity scores. Despite partial improvements through $log(x+1)$ normalization and severity inclusion, SARIMAX proved unreliable in practice due to data sparsity, short time horizons, and high burstiness typical of real-world vulnerability activity. Confidence intervals frequently expanded unrealistically, and negative forecasts occasionally appeared. To address these limitations, we explored count-based models (Poisson regression) which better capture the discrete and event-driven nature of sightings data. Early results show that these models provide more stable and interpretable forecasts, particularly when data are aggregated weekly.

We also discuss alternative strategies better suited for operational settings, including exponential decay functions for short-lived forecasting horizons to estimate future activity levels without relying on long historical series. This study highlights both the promise and the inherent challenges of forecasting rare, bursty cyber events, providing methodological insights toward integrating predictive analytics into vulnerability intelligence workflows.
\end{abstract}

\vspace{1em}
\noindent\textbf{Topics:} \textit{Measuring vulnerabilities}, \textit{Forecast}, \textit{Decision}

\section{Introduction}
Vulnerability sightings are reports or observations that a specific vulnerability is being referenced, discussed, or even exploited in the wild \cite{bonhomme2025scoring}. Such sightings (e.g., exploit proofs-of-concept, scanner detections, or blog references) provide concrete evidence that a vulnerability is active. Tracking and forecasting sightings can help defenders prioritize patches and assess risk based on real-world activity. We also leverage our earlier VLAI severity model (Bonhomme \& Dulaunoy 2025 in \cite{bonhomme2025vlairobertabasedmodelautomated}) which predicts CVSS-like scores from text descriptions. In particular, the VLAI “severity score” can serve as an exogenous indicator of a vulnerability’s potential impact. In this paper, we present experiments on forecasting the future number of sightings for individual vulnerabilities using statistical models. We compare SARIMAX time-series models and Poisson regression approaches, discuss their challenges, and propose alternative strategies and future directions.

\section{Related Work}
The forecasting of vulnerability-related activities remains a relatively unexplored area within cyber threat intelligence research. Most prior work has focused on the assessment rather than the prediction of vulnerability risk. Traditional models rely on CVSS scoring systems or manually curated indicators, which, while informative, fail to capture the dynamic nature of vulnerability exploitation and community attention.

Our previous work, VLAI, introduced a transformer-based model designed to predict software vulnerability severity directly from textual descriptions. Trained on over 600,000 real-world entries, VLAI provides consistent and automated severity scoring, offering a robust foundation for downstream analytical tasks. In the present study, we extend this line of research by integrating the VLAI severity score as an explanatory variable for forecasting future sightings of vulnerabilities, thus linking semantic severity assessment with temporal activity prediction. We are still regularly maintaining and improving this model.

In parallel, the Vuln4Cast project (FIRST, 2022, \cite{10.1145/3492328}) proposed a data-driven framework for forecasting vulnerability publication. While Vuln4Cast emphasizes global trends across large vulnerability datasets, our work operates at a finer granularity, focusing on individual CVEs and their observable evolution across open intelligence sources.

The Exploit Prediction Scoring System (EPSS) presented in \cite{jacobs2023enhancingvulnerabilityprioritizationdatadriven} provides another relevant line of work, aiming to estimate the probability of future exploitation for a given vulnerability based on historical exploitation patterns and vulnerability metadata. While EPSS and our approach share the goal of improving vulnerability prioritization, the focus here is different: rather than estimating the likelihood of exploitation, we aim to forecast observable activity over time (e.g., sightings of PoCs, detection scanner templates, or related discussions). Our models therefore operate on temporal event data, seeking to understand and anticipate vulnerability attention dynamics rather than discrete exploitation outcomes.

Prior research on time-series modeling in cybersecurity has primarily addressed incident frequency or malware propagation, rarely addressing sparse event data such as vulnerability sightings. By adapting statistical and probabilistic models to this context, our work contributes a new perspective at the intersection of vulnerability intelligence, predictive analytics, and cyber situational awareness.

\section{SARIMAX Time-Series Forecasting}
We first applied classical ARIMA-family models to the daily sighting counts of a given vulnerability. The data are very sparse (often mostly zeros with occasional small counts), which complicates modeling.

\subsection{Seasonal SARIMAX}
We attempted SARIMAX model including seasonal terms, with the daily count series log-transformed. However, this yielded poor results: the fitted model produced nonsensical forecasts (sometimes negative or near-zero far into the future before the logorithmic transformation) and extremely wide confidence intervals (CI). This is unsurprising given the data characteristics: SARIMAX assumes a relatively smooth stationary series with clear autoregressive or seasonal structure, but our sighting data are low-count and burst-like. As a result, the model over-smooths the few spikes and extrapolates flat or negative trends. Practically, forecasting sighting counts with ARIMA models often requires many more data points (on the order of dozens or more) than we have. In short time series ($<$ 30 days of data), SARIMAX is unreliable: it struggles to fit parameters and produces unstable forecasts.

%

\subsection{SARIMAX without seasonal components}
We simplified the approach by omitting seasonal components and using a log(x+1) transform on counts. In addition, we included the VLAI severity score as an exogenous regressor (exog) to capture whether higher-severity vulnerabilities tend to generate more sightings. The VLAI parameter is generally stable over time but may vary following regular VLAI Severity model
updates\footnote{\scriptsize \url{https://doi.org/10.57967/hf/5926}}.
We are currently able to re-train the model in approximately 7 hours\footnote{\scriptsize \url{https://github.com/vulnerability-lookup/VulnTrain}}.\\
The practical workflow was:

\begin{itemize}
    \item Aggregate daily sightings: Sum all reported sightings for the CVE across sources (e.g. exploit databases, Fediverse posts) per day.
    \item Compute daily VLAI severity: Take the vulnerability’s VLAI score each day (often nearly constant after publication).
    \item Use the daily sighting count (log-transformed) as the target and the daily severity score as an exogenous covariate.
    \item Predict future counts by feeding the model the average recent severity as the assumed exogenous value (since severity changes little day-to-day).
\end{itemize}

This model effectively tests whether severity influences sighting volume. The log(x+1) transform stabilized variance and prevented fitting errors on zeros. We found marginal improvements: the forecasts were more sensible than in the first experience with seasonal components, and peak events were captured slightly better.

However, significant issues remained. With only 10–15 days of training data, a SARIMAX model can overfit or produce negative predictions. In practice, we observed that having around 10 days of sightings for a vulnerability is \textbf{already valuable for our analysis}, but unfortunately, it is insufficient for reliable SARIMAX forecasting.
Sudden spikes of sightings in the training set caused the model to extrapolate downward into negative territory afterwards.
The confidence interval of the forecast often spans many orders of magnitude, reflecting the model's high uncertainty given limited data. In practice, ARIMA-based methods usually require on the order of 50–100 observations for reliable estimation.
In this case, the results can sometimes improve, but requiring such extensive data is impractical for our purposes. SARIMAX can, however, perform well when forecasting the overall evolution of vulnerabilities, as demonstrated in \cite{10.1145/3492328}.

\section{Poisson Regression for Count Forecasting}

Given the limitations of SARIMAX on sparse counts, we explored count-based regression models. We treated the number of daily (or weekly) sightings as a Poisson outcome.
Poisson regression is designed for non-negative integer data. It naturally ensures that forecasts remain $\geq 0$, unlike linear models. It can also include covariates (e.g. days since disclosure, severity) and account for overdispersion via a negative binomial variant if needed.

Initial results showed Poisson forecasts more plausibly non-negative. However, like any regression, it still requires enough data. In practice, we sometimes saw under-dispersion (variance < mean) or over-dispersion (variance > mean) which violates the basic Poisson assumption.

Poisson models handle the discrete nature of sightings well, but still struggle with data scarcity and variability. They can complement time-series methods: for example, one might fit a Poisson model to the first few weeks after disclosure and use it for short-term prediction, then switch to another model as more data accumulates.

We applied this technique to the top vulnerabilities—those with a high number of sightings—reported in our monthly vulnerability reports\footnote{\scriptsize \url{https://www.vulnerability-lookup.org/tags/vulnerabilityreport}}
and observed improved results compared to ARIMA-based methods. The outcomes indicate general trends, but should not be interpreted as precise predictions.

\subsection{Issues and Limitations}

From the previous experiments based on the Poisson regression, several recurring issues emerged:

\begin{itemize}
 \item Very short series: With only 10–30 days of data per CVE, most statistical models are at their limits. For ARIMA/SARIMAX, estimating even a basic model needs on the order of 50 observations (which is a very good number for us) to constrain parameters. Insufficient data leads to overfitting or unconstrained predictions. We recommend collecting longer histories (weeks or months) before relying on ARIMA forecasting.
 \item Spikes and outliers: Sudden jumps in sightings (e.g. a major exploit release) dominate the fit. ARIMA tries to treat these as trend, resulting in unrealistic subsequent behavior.
 \item Negative or nonsensical forecasts: Linear extrapolation can drive counts below zero or to implausible values. This is especially problematic for count data. Using Poisson (or other non-negative) models addresses this.
 \item Exploding confidence intervals: When a model is highly uncertain (due to short series or high variance), the predicted range can become absurdly large. This indicates the model “knows” it has too little information.
 \item Exogenous variables: Including severity (VLAI or CVSS) as an exogenous regressor is sensible in theory (higher-severity flaws may draw more attention). However, because the severity score itself is nearly constant immediately after disclosure, it provides limited variation. In our SARIMAX fits, the severity coefficient was hard to estimate reliably until many observations accumulated.
\end{itemize}

\section{Exponential Decay Model for short-term forecasting}

The function for the decaying exponential is given by:

\begin{equation}
y(t) = a \cdot e^{-bt} + c
\end{equation}

This model is appropriate for vulnerabilities already past their peak.
If a vulnerability is still rising (no decay yet), the fit might look flat or strange.

\begin{lstlisting}[style=Python, caption=Fitting the exponential decay model]
def exp_decay(t, a, b, c):
    return a * np.exp(-b * t) + c

t = np.arange(len(daily_series))
y = daily_series.values

popt, pcov = curve_fit(exp_decay, t, y, p0=(max(y), 0.5, min(y)))
a, b, c = popt
\end{lstlisting}

\section{Logistic Growth}
The function for the logistic growth is given by:

\begin{equation}
y(t) = \frac{L}{1 + e^{-k(t-t_0)}}
\end{equation}

where:
\begin{itemize}
    \item L: Maximum expected number of sightings (plateau)
    \item k: Growth rate
    \item $t_0$: Day of inflection (midpoint of growth)
\end{itemize}

This model is relevant for “burst-and-fade” dynamics such as CVE mentions, social signals, etc.
This model is more fitted for newly published or trending vulnerabilities.
It captures what we typically observe: a rapid increase in reports just after disclosure,
then a plateau and slow decay as attention fades.

\begin{lstlisting}[style=Python, caption=Fitting the logistic model with initial guess for parameters]
def logistic(t, L, k, t0):
    return L / (1 + np.exp(-k * (t - t0)))

L0 = max(y) * 1.5  # estimated upper bound
k0 = 0.3           # moderate growth rate
t0 = np.median(t)  # inflection around middle

popt, _ = curve_fit(
    logistic, t, y, p0=[L0, k0, t0],
    bounds=([0, 0, 0], [np.inf, np.inf, np.inf]),
    maxfev=10000
)

last_date = daily_series.index.max()
t_future = np.arange(len(daily_series), len(daily_series) + future_steps)  # days after last observation

forecast_values = np.maximum(logistic(t_future, L, k, t0), 0)
\end{lstlisting}

\texttt{curve\_fit} bounds ensures sensible and non-negative parameters (L, k, $t_0$).

\section{Adaptive Logistic or Exponential Decay Forecast}

Another alternative is to detect based on the sigthings repartition which strategy is the best.

\begin{itemize}
 \item Trend detection: Checks if recent sightings are increasing or decreasing (via linear slope) 
 \item Model selection: Uses logistic if slope is positive, else exponential decay.
 \item Curve fitting: To estimate parameters.
 \item Forecasting: Extends the model 10 days into the future.
\end{itemize}

%
%

\section{Experiences}

This section presents a few experiments with the different models, illustrating and supporting our conclusions. We used both recent vulnerabilities and vulnerabilities with a long history of recorded sightings (often exploited vulnerabilities detected via the Shadow server).
The experiments can be reproduced using the project code at commit \texttt{8678ab6a4e7a036eb578f1310b42aa3a22c686ea}\footnote{\scriptsize \url{https://github.com/vulnerability-lookup/TARDISsight/tree/8678ab6a4e7a036eb578f1310b42aa3a22c686ea}}

\subsection{CVE-2025-61932}

\begin{figure}[H]
    \centering
    \includegraphics[width=\linewidth]{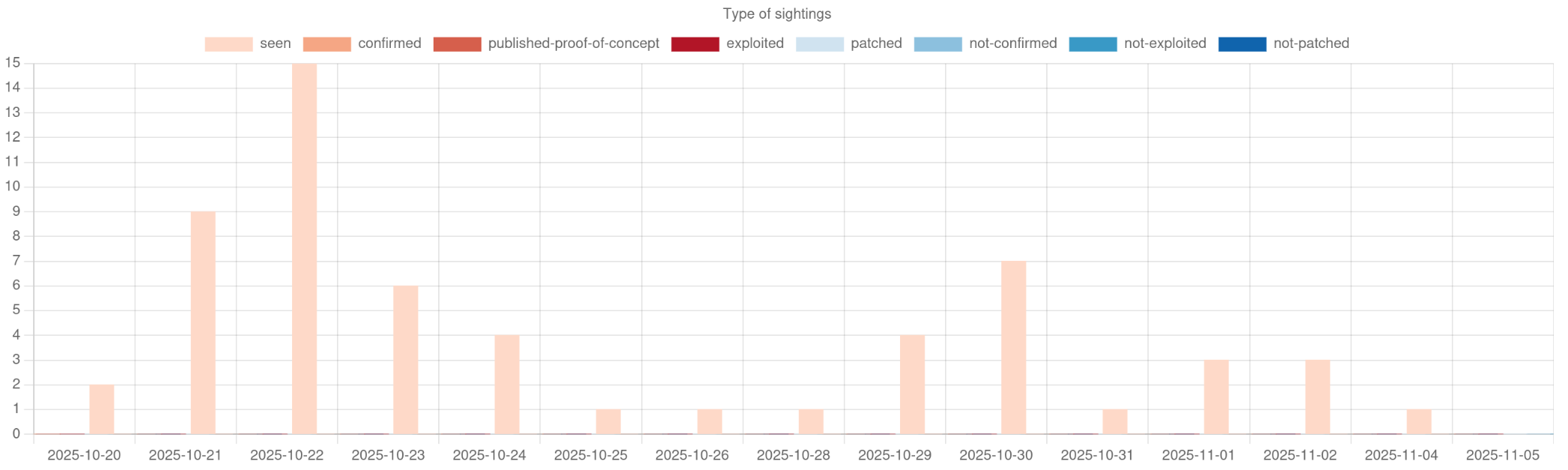}
    \caption{Observed sightings over time for CVE-2025-61932}
\end{figure}

\begin{figure}[H]
    \centering
    \includegraphics[width=\linewidth]{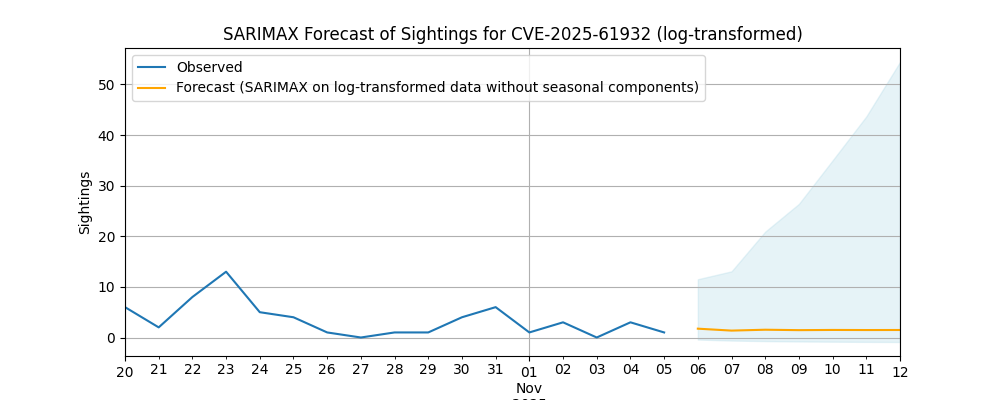}
    \caption{SARIMAX with Log-transform counts without seasonal components}
    \label{fig:CVE-2025-61932-SARIMAX}
\end{figure}

Other tests produced results similar to \cref{fig:CVE-2025-61932-SARIMAX}, with notably wide confidence intervals.
Without the logarithmic transformation, the forecasted sighting counts can occasionally become negative.

\begin{figure}[H]
    \centering
    \includegraphics[width=\linewidth]{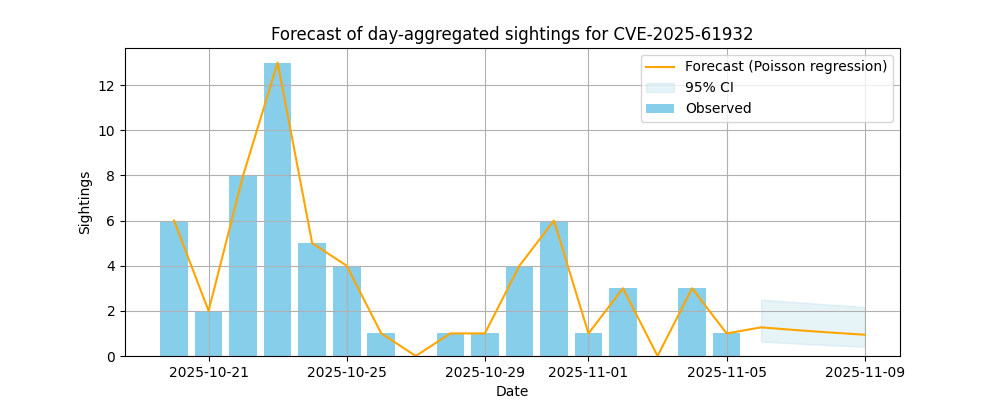}
    \caption{Poisson regression}
     \label{fig:CVE-2025-61932-POISSON}
\end{figure}

\begin{figure}[H]
    \centering
    \includegraphics[width=\linewidth]{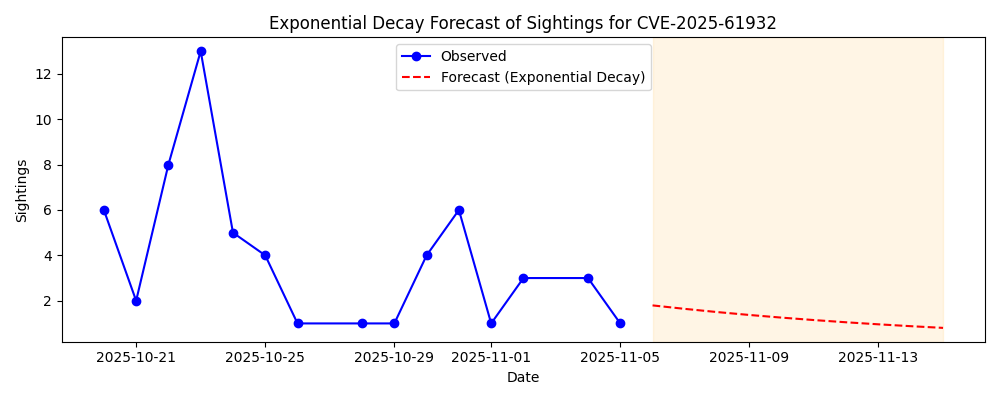}
    \caption{Exponential decay}
    \label{fig:CVE-2025-61932-Exponential}
\end{figure}

With a sufficient number of sightings, the Poisson regression typically produces results comparable to the exponential decay method, as illustrated in \cref{fig:CVE-2025-61932-POISSON} and \cref{fig:CVE-2025-61932-Exponential}.

\begin{figure}[H]
    \centering
    \includegraphics[width=\linewidth]{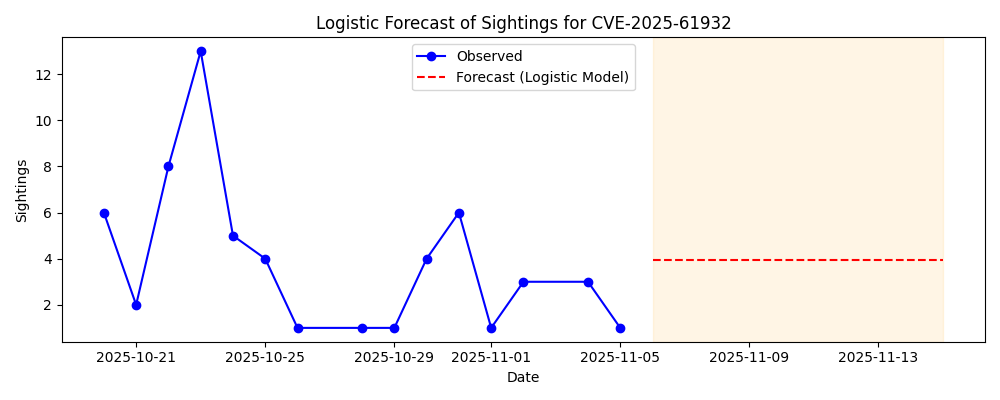}
    \caption{Logistic model}
\end{figure}

\begin{figure}[H]
    \centering
    \includegraphics[width=\linewidth]{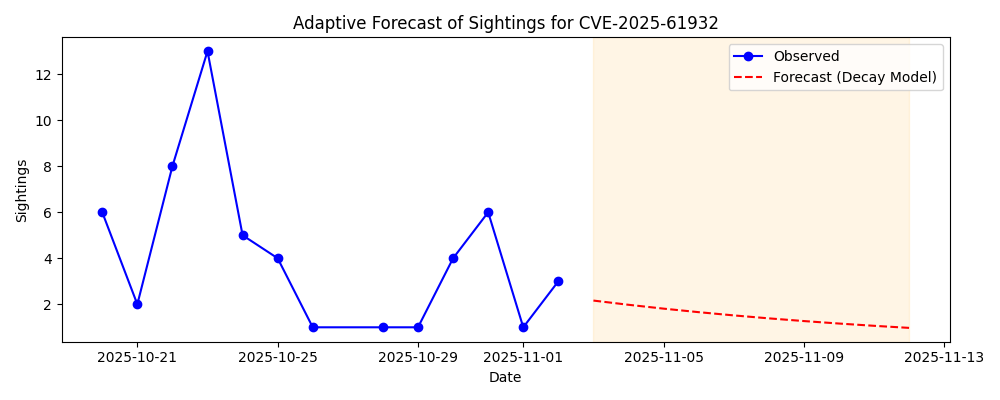}
    \caption{Logistic model with sigthings up to 2025-11-01}
    \label{fig:CVE-2025-61932-adaptive-simlation-forecast}
\end{figure}

The \cref{fig:CVE-2025-61932-adaptive-simlation-forecast} was produced with the same logistic model, excluding all observations of the vulnerability after 2025-11-01, in order to assess the accuracy of its forecast.

\subsection{CVE-2025-59287}

\begin{figure}[H]
    \centering
    \includegraphics[width=\linewidth]{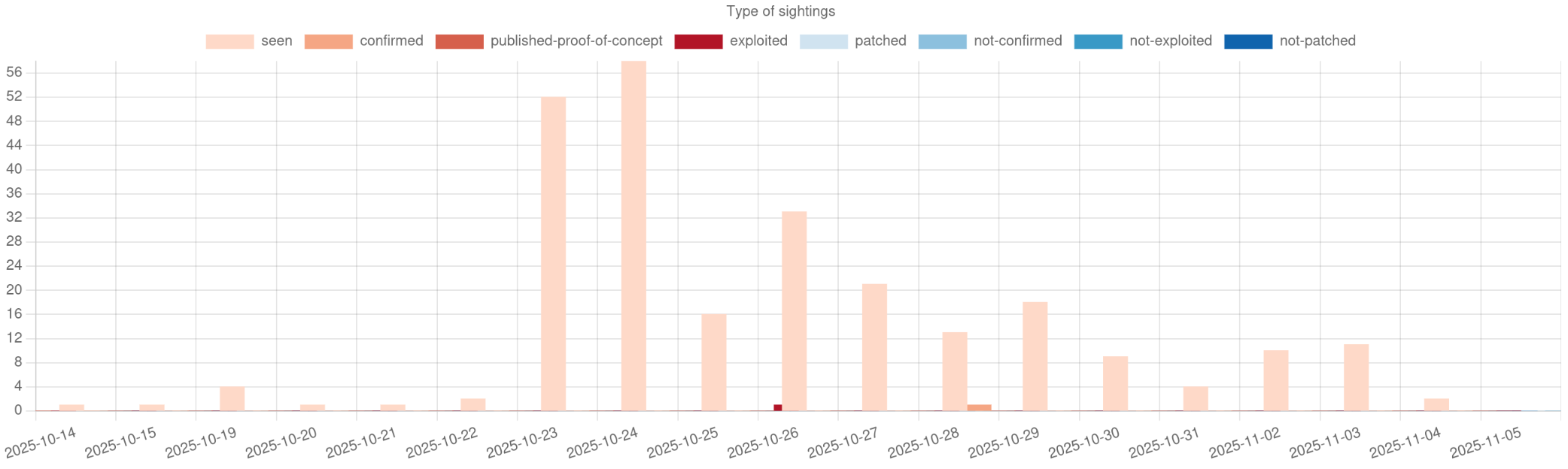}
    \caption{Observed sightings over time for CVE-2025-59287}
\end{figure}

\begin{figure}[H]
    \centering
    \includegraphics[width=\linewidth]{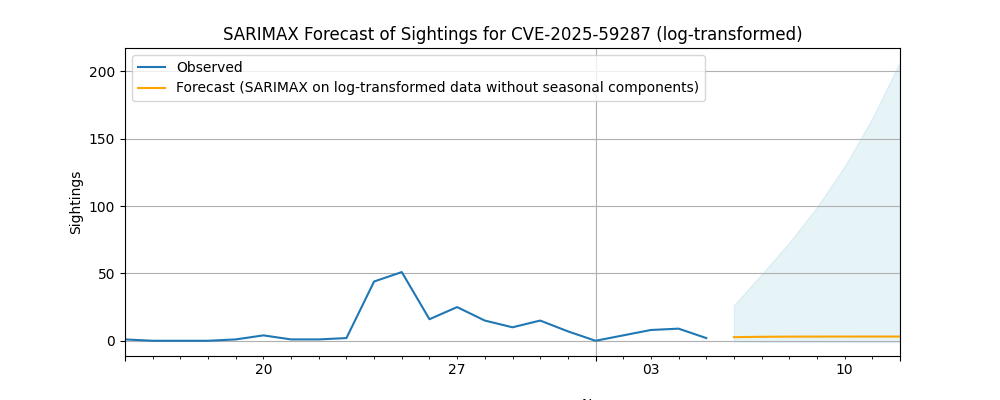}
    \caption{SARIMAX with Log-transform counts without seasonal components}
     \label{fig:CVE-2025-59287-SARIMAX}
\end{figure}

Once again, the SARIMAX forecast (\cref{fig:CVE-2025-59287-SARIMAX}) is largely unsuitable for our case,
even with the logarithmic transformation.
The confidence intervals remain very wide, regardless of the distribution of sightings.

\begin{figure}[H]
    \centering
    \includegraphics[width=\linewidth]{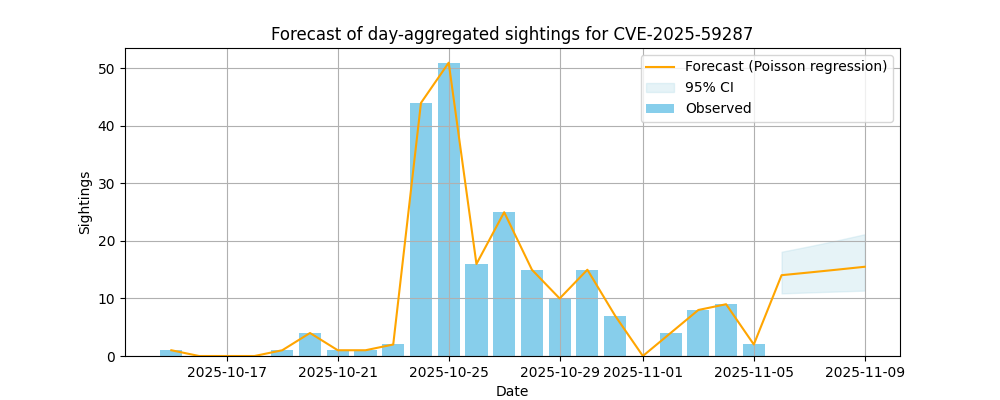}
    \caption{Poisson regression}
    \label{fig:CVE-2025-59287-poisson-forecast}
\end{figure}

\begin{figure}[H]
    \centering
    \includegraphics[width=\linewidth]{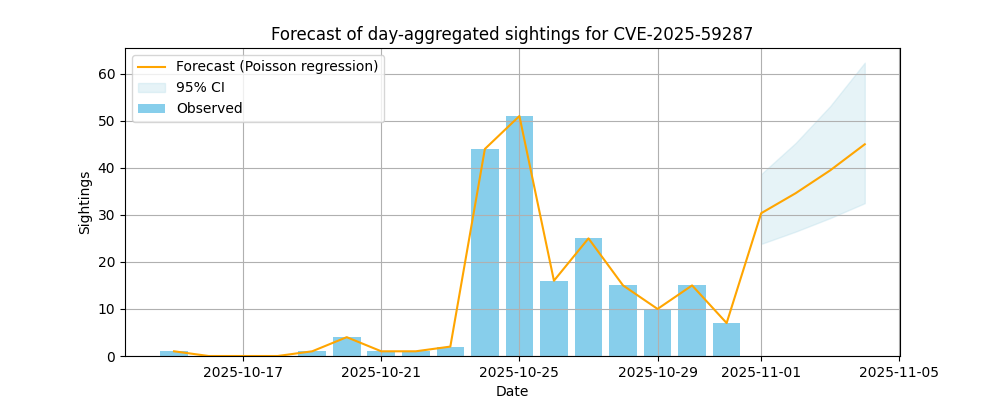}
    \caption{Poisson regression}
    \label{fig:CVE-2025-59287-simulation-forecast}
\end{figure}

The figure \cref{fig:CVE-2025-59287-simulation-forecast} was generated using the same Poisson model as \cref{fig:CVE-2025-59287-poisson-forecast}, but with all observations after 2025-11-01 removed to evaluate the forecast’s accuracy. We observe that the growth appears stronger in \cref{fig:CVE-2025-59287-simulation-forecast}, which is expected, but the final prediction overestimates the reality.

\begin{figure}[H]
    \centering
    \includegraphics[width=\linewidth]{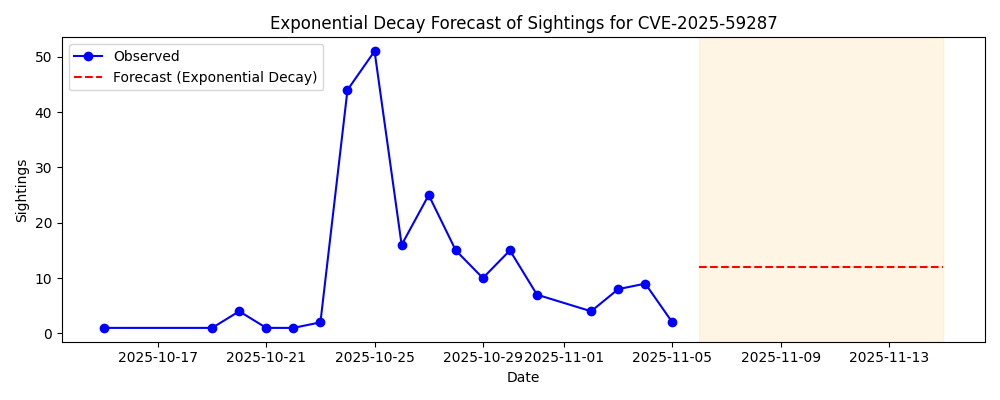}
    \caption{Exponential decay}
    \label{fig:CVE-2025-59287-decay}
\end{figure}

Another observation is that a sudden drop in sightings collection is often not visible with the Poisson regression. As explained in the previous section, a practical solution is to detect which model to use based on the linear slope (analysis of recent trends). In this case the exponential decay would have made more sense (\cref{fig:CVE-2025-59287-decay}).

The \textbf{adaptive solution} would have selected the exponential decay model.\footnote{\tiny \url{https://github.com/vulnerability-lookup/TARDISsight/blob/8678ab6a4e7a036eb578f1310b42aa3a22c686ea/tardissight/decay/adaptive.py\#L75}}

%
%


\subsection{CVE-2022-26134}

\begin{figure}[H]
    \centering
    \includegraphics[width=\linewidth]{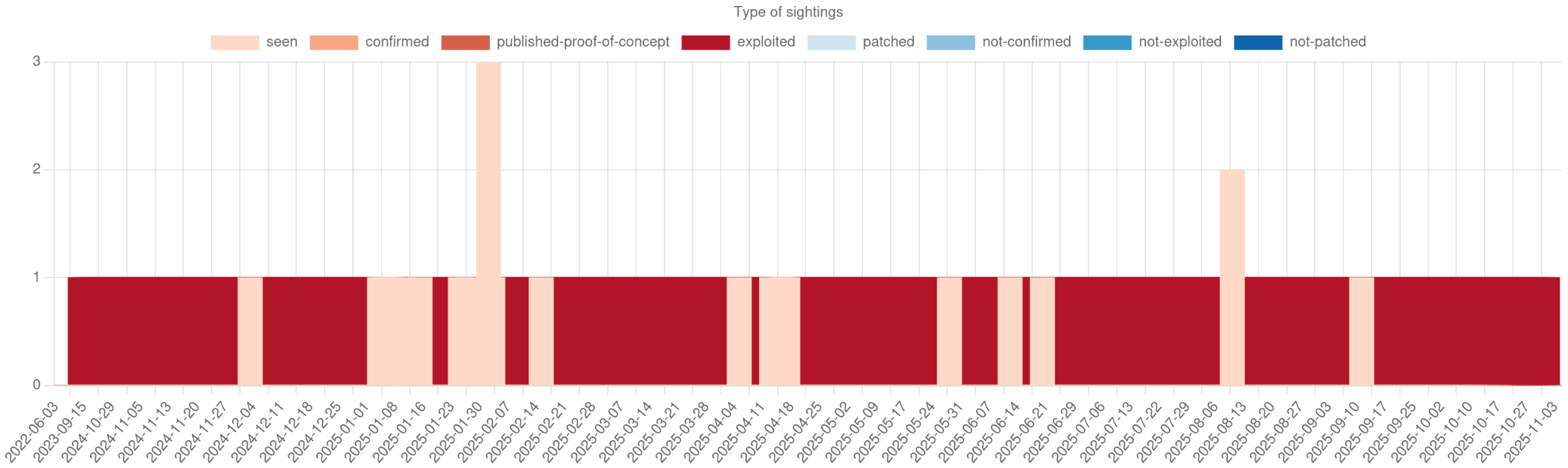}
    \caption{Observed sightings over time for CVE-2022-26134}
\end{figure}

\begin{figure}[H]
    \centering
    \includegraphics[width=\linewidth]{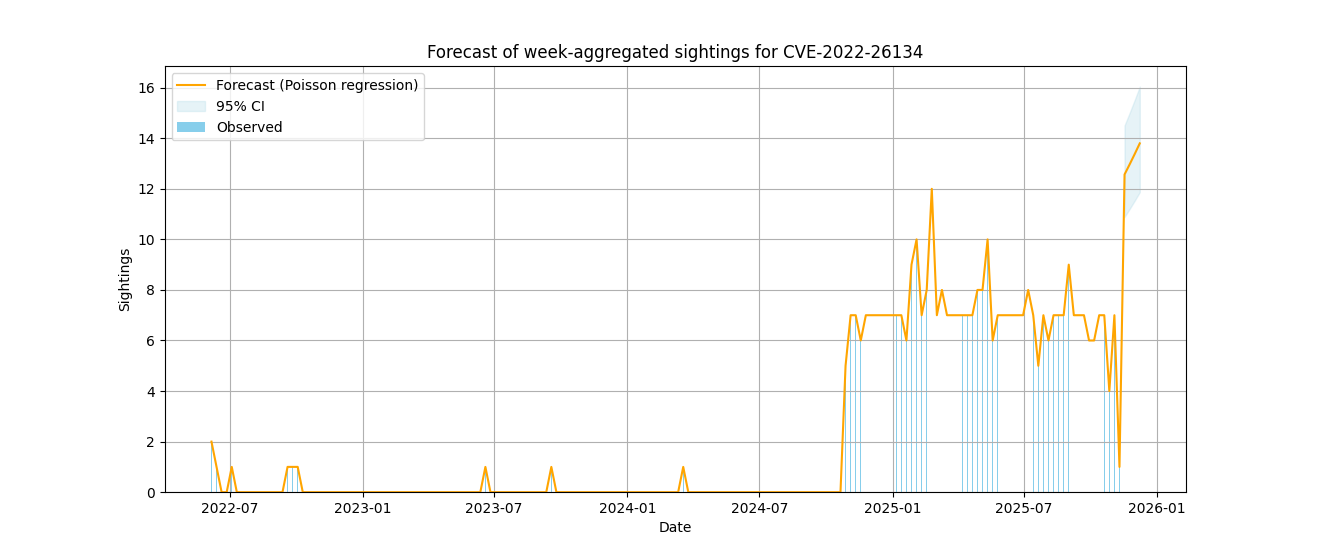}
    \caption{Poisson regression}
\end{figure}

\begin{figure}[H]
    \centering
    \includegraphics[width=\linewidth]{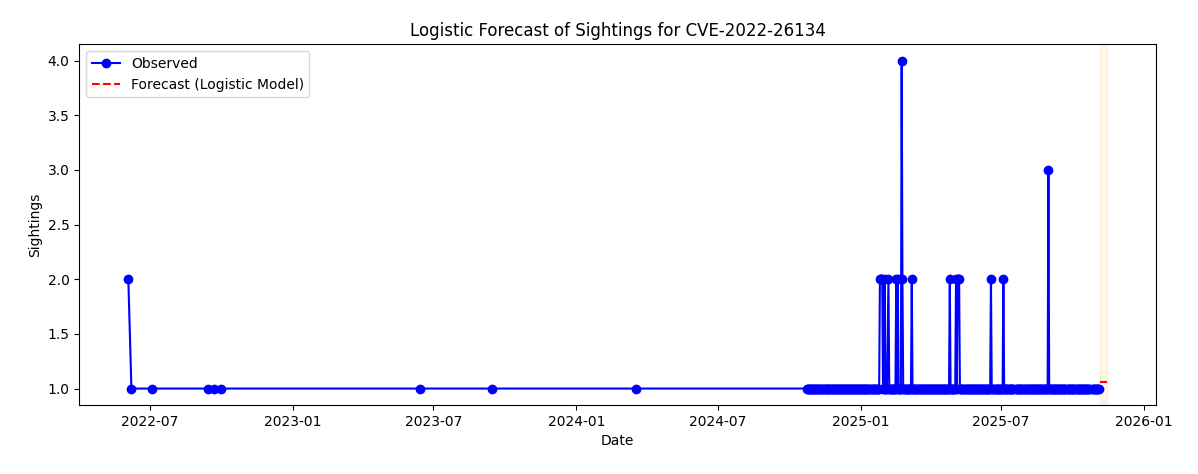}
    \caption{Logistic model}
\end{figure}

While the decay model proves inadequate here, the logistic model appears largely unaffected by recent bursts of sightings for vulnerabilities monitored over an extended period (several months or years) with regular observations.
We confirmed this behavior using sightings primarily sourced from the Shadowserver project.

\section{Practical suggestions}

Based on our experience, we suggest:

\begin{itemize}
 \item Prioritize gathering more data per vulnerability (e.g. combine multiple sources, extend the observation window).
 \item Use simple models initially. For very short windows, even a rolling average or exponential decay will outperform SARIMAX.
 \item Check for data quality issues (missing days, duplicate reports, extreme spikes) and handle them before modeling (clamping, logarithmic transformation).
 \item If negative forecasts appear, switch to a count-based approach like Poisson.
 \item Detecting bursts of sightings immediately after a vulnerability’s publication can help in selecting a more appropriate model and in setting the initial growth rate and upper bound for the logistic approach.
 \item Reserve SARIMAX with seasonality only for much longer series (e.g. if a vulnerability activity remains visible for months). We need to test more in the future SARIMAX for vulnerabilities with observations from the Shadowser foundation.
\end{itemize}

\section{Future work}
Real-time updating. In a production system (such as Vulnerability-Lookup), a forecasting module could update predictions daily as new sightings arrive. This would allow forecasts to be adjusted on the fly.\\
Detecting unusual spikes and missing days of sightings to select the most appropriate model.\\
Currently, all sightings are treated as equivalent, regardless of their type (Seen, Confirmed, Exploited, etc.). We aim to incorporate both the different types of sightings and the severity scores from our VLAI model. It could help us establish a link with the actual exploitation of a vulnerability.

\bibliography{mybib}{}
\bibliographystyle{plain}
\end{document}